\documentclass[11pt]{article}
\usepackage[T1]{fontenc}
\usepackage[margin=1in]{geometry}
\usepackage{microtype}
\usepackage{amsmath,amssymb}
\usepackage{graphicx}
\usepackage{subcaption}
\usepackage{caption}
\usepackage{booktabs}
\usepackage{siunitx}
\usepackage[hidelinks]{hyperref}     
\usepackage{xurl} 
\usepackage[nameinlink,capitalise]{cleveref}
\usepackage{xcolor}
\usepackage{longtable}
\usepackage{float}
\usepackage{placeins}

\definecolor{EMO}{HTML}{E69F00}
\definecolor{CSF}{HTML}{56B4E9}
\definecolor{SPF}{HTML}{009E73}
\definecolor{STR}{HTML}{CC79A7}

\graphicspath{{figs/}}

\title{Module control in youth symptom networks across COVID-19}
\author{Tianyi Fan$^{2}$ \and Xizhe Zhang$^{1,2,*}$}
\date{%
    $^1$ Early Intervention Unit, Department of Psychiatry, The Affiliated Brain Hospital of Nanjing Medical University, Nanjing, China\\%
    $^2$ School of Biomedical Engineering and Informatics, Nanjing Medical University, Nanjing, China\\[0.5ex]%
    $^{*}$ Corresponding author: zhangxizhe@gmail.com%
}

\begin{document}

\maketitle

\begin{abstract}
The COVID-19 pandemic exposed young people to a prolonged and evolving societal stressor, yet it remains unclear whether symptom networks were reorganized or whether control was redistributed across a conserved modular scaffold. Here we analysed repeated cross-sectional data on 47 self-reported mental-health symptoms from 14,181 U.S. young adults aged 18–24 years across five COVID-19 phases between 2020 and 2023. For each phase, we estimated Gaussian graphical models, identified symptom communities, and characterized minimum-dominating-set-based module control. Symptom networks showed broadly conserved community organization across phases, indicating a stable mesoscale scaffold despite marked temporal variation. By contrast, intermodule control shifted from an early configuration centered on stress-related symptoms to a later, more distributed pattern spanning emotional, cognitive and social domains. Resampling analyses showed high stability for node strength and moderate stability for module-to-module control, whereas average within-module control was less robust. These findings suggest that prolonged crisis may preserve the modular architecture of youth psychopathology while redistributing control across symptom domains, and they identify intermodule control as a comparatively robust mesoscale feature for cross-phase comparison.
\end{abstract}

\section{Introduction}\label{sec:introduction}

The COVID-19 pandemic created a prolonged, population-wide stressor with substantial mental health consequences, particularly for adolescents and young adults~\cite{xiong_impact_2020,holmes_multidisciplinary_2020,covid-19_mental_disorders_collaborators_global_2021,guessoum_adolescent_2020,panchal_impact_2023,kalisch_resilience_2017,zheng_psychiatry_2025}. Disruptions in education, social networks, and economic security, together with sustained uncertainty and health threat, were accompanied by marked increases in anxiety, depressive symptoms, and stress-related complaints in youth across multiple countries~\cite{racine_global_2021,loades_rapid_2020,pierce_mental_2020,singh_impact_2020,viner_school_2022}. At the same time, the pandemic did not unfold as a single homogeneous event: waves of infection, changing variants, and evolving non-pharmaceutical interventions (NPIs) such as school closures, stay-at-home orders, and mobility restrictions produced a sequence of qualitatively distinct phases~\cite{viner_school_2022,hale_global_2021}. In the United States, daily case counts and policy stringency indicators provide a concise macro-level depiction of this staged trajectory, characterized by sharp early surges under stringent restrictions, followed by repeated waves and an eventual transition toward lower average stringency~\cite{hale_global_2021}. Against this backdrop, a central scientific challenge is to understand not only whether youth psychopathology burden increased, but how the internal organization of symptoms adapted as epidemiological pressure and policy regimes shifted over time.

Network psychometrics conceptualizes mental disorders as systems of mutually reinforcing symptoms rather than manifestations of a single latent disease entity~\cite{borsboom_network_2013,borsboom_network_2017}. In this framework, nodes represent specific symptoms and edges represent conditional dependencies or putative causal influences among them, motivating graph-theoretic tools to identify influential symptoms and to characterize how symptom networks differ across individuals, groups, and contexts~\cite{epskamp_estimating_2018,zhang_control_2022,pan_connectivity_2024,pan_adaptive_2026}. Network control theory extends this systems perspective by asking which elements of a network are capable, in principle, of steering the system's state toward or away from dysfunction~\cite{liu_controllability_2011,gu_controllability_2015}. While early applications emphasized node-level controllability metrics, there is growing recognition that clusters of tightly connected symptoms---modules---may function as higher-order control units that better match the scales at which psychological and social interventions are implemented~\cite{pan_module_2024}.

Building on this idea, the module control framework defines control using minimum dominating sets (MDSs)~\cite{haynes_fundamentals_2013,zhao_statistical_2015,pan_module_2024}. An MDS is a smallest set of ``driver'' nodes such that every node in the graph is either a driver or a neighbor of at least one driver~\cite{haynes_fundamentals_2013}. By enumerating all minimum-cardinality MDSs, one can quantify how frequently each symptom appears in a driver set (node control frequency) and, by aggregating across communities, derive two complementary indices of module-level control: the average control frequency (ACF), which summarizes normalized driver participation of nodes within a module, and the average module control strength (AMCS), which quantifies how strongly a source module's drivers dominate nodes in target modules. Together, these indices define a directed module control network (MCN) that captures how influence is distributed across modules at the mesoscale~\cite{pan_module_2024}.

Despite rapid methodological progress, most empirical demonstrations of module-level control have been cross-sectional and single-wave~\cite{pan_module_2024}. We therefore know comparatively little about how module organization and controllability evolve under a sustained, society-wide stressor such as the COVID-19 pandemic. Several questions remain open. First, does the modular scaffold of youth symptom networks remain stable as the pandemic unfolds, or does the crisis fundamentally reorganize community structure? Second, even if a mesoscale scaffold persists, does the distribution of control across modules reconfigure over time---for example, from an early regime dominated by stress-related circuitry to a later regime in which multiple domains jointly govern network dynamics? Third, are module-level control metrics sufficiently robust to sampling variability and case loss to support comparisons across phases or cohorts? These questions are especially salient because community detection has known resolution and near-degeneracy issues~\cite{fortunato_resolution_2007,fortunato_community_2010,newman_finding_2004,newman_modularity_2006,reichardt_statistical_2006}, and network metrics require explicit resampling-based evaluation of accuracy and stability before they can be used as biomarkers or intervention guides~\cite{epskamp_estimating_2018,efron_bootstrap_1979,efron_introduction_1994}.

Moreover, existing work has rarely embedded symptom networks within an explicit macro-level description of the pandemic. From a systems perspective, youth psychopathology can be viewed as a micro-level subsystem evolving within a time-varying macro context defined jointly by epidemiological pressure and policy responses~\cite{holmes_multidisciplinary_2020}. Daily infection rates and NPIs determine both the objective risk environment and the degree of disruption to schooling, mobility, and social contact~\cite{viner_school_2022,hale_global_2021}. These macro-level trajectories can be summarized by time series of case counts and policy stringency indices, which exhibit clear phase structure in the United States~\cite{hale_global_2021,su_temporal_2024,zhou_functional_2024}. Conceptually, this suggests a dual-timescale picture: on a fast timescale, individual symptoms fluctuate and interact; on a slower timescale, changes in macro-level conditions may gradually reweight the relative influence of symptom modules without necessarily dismantling the underlying modular scaffold. Empirically linking macro shocks to micro-level module control remains an important but understudied problem.

In the present study, we apply a module-control framework to repeated cross-sectional data from U.S. young adults aged 18–24 years (N = 14,181) who completed the 47-item Mental Health Quotient between 2020 and 2023 \cite{newson_assessment_2022,newson_mhq_2024,bala_hierarchy_2024}. We stratify responses into five COVID-19 phases and, within each phase,  identify symptom communities and quantify domination-based control through exact enumeration of all minimum dominating sets. This design allows us to address three questions. First, does the mesoscale community scaffold of youth symptom networks remain broadly conserved across pandemic phases? Second, against this scaffold, is intermodule control redistributed over time, shifting from a more stress-centered configuration to a more distributed cross-domain pattern? Third, which network metrics are sufficiently robust to support cross-phase comparison? By focusing on structural stability, intermodule control reallocation, and metric robustness, we aim to characterize how prolonged crisis is reflected in the mesoscale organization of youth symptom networks and to identify the control-related quantities most suitable for repeated cross-sectional comparison.

\section{Methods}\label{sec:methods}

\subsection{Study design and sample}

This study used phase-stratified repeated cross-sectional data from the Global Mind Project (GMP), an ongoing online assessment of population mental health and wellbeing. We focused on U.S. young adults aged 18--24 years who completed the 47-item Mental Health Quotient (MHQ) during the first four years of the COVID-19 pandemic. The macro-level contextual time series covered calendar weeks from 2020-01-01 to 2023-12-31. The integrated GMP export used for phase-stratified network estimation was provided in a pre-cleaned form and contained eligible records from 2020-04-14 onward.

To align the analytic sample with major infection waves and shifts in non-pharmaceutical interventions, we partitioned the data into five calendar-based pandemic phases. Phase-specific sample sizes are reported in Table~\ref{tab:phase_samples}. Across phases, the final U.S. analytic sample comprised 14{,}181 respondents. These phase-specific samples were treated as repeated cross-sections from the same source population rather than as a longitudinal cohort.

\begin{table}[ht]
\centering
\caption{Pandemic phase definitions and sample sizes (U.S. cohort).}
\label{tab:phase_samples}
\begin{tabular}{l l l r}
\toprule
\textbf{Phase} & \textbf{Start} & \textbf{End} & \textbf{$n$} \\
\midrule
Early        & 2020-04-14 & 2020-07-01 & 518  \\
First Wave   & 2020-07-01 & 2021-01-01 & 2321 \\
Second Wave  & 2021-01-01 & 2022-01-01 & 4721 \\
Omicron      & 2022-01-01 & 2023-01-01 & 4023 \\
Post-Omicron & 2023-01-02 & 2024-01-01 & 2598 \\
\bottomrule
\end{tabular}

\vspace{0.3em}
\parbox{0.92\linewidth}{\footnotesize \textit{Note.} The integrated GMP export used here was provided in a pre-cleaned form following project-level quality filters. Because item-level exclusion counts were not retained in the integrated export, we report final analytic sample sizes per phase after applying the preprocessing pipeline described in Section~2.2.}
\end{table}

\textbf{External replication cohort (India).}
To evaluate whether the mesoscale scaffold and control redistribution patterns were specific to the U.S. sample, we constructed an independent replication cohort from India using the same GMP platform and MHQ instrument. We restricted records to respondents aged 18--24 years with response dates within the same calendar window used for phase labeling (2020-04-14 to 2024-01-01) and assigned phases using the identical cut points in Table~\ref{tab:phase_samples}. The resulting India cohort comprised 31{,}626 respondents (phase-specific $n$: Early 345; First Wave 5{,}366; Second Wave 8{,}942; Omicron 8{,}757; Post-Omicron 8{,}216). All network estimation, community detection, and control analyses were then re-run on India phase strata using the same parameters and preprocessing rules as in the U.S. cohort.

\textbf{Sample-size matching sensitivity analysis.}
Because the India phase samples were larger than the U.S. samples, we additionally performed a phase-wise downsampling sensitivity analysis to control for potential sample-size effects on network stability and downstream control quantities. For each phase, we repeatedly subsampled the India cohort without replacement to match the corresponding U.S. phase sample size, re-estimated the network and communities, and recomputed module- and domain-level control summaries. Downsampling results are reported alongside the full-sample India estimates in the Supplement.

\textbf{Demographic characterization and variable availability.}
Demographic variables available in the GMP export (including sex, education, employment status, and household income) were used to characterize cohort composition. Several fields were introduced or reformatted across survey versions and therefore exhibit phase-dependent missingness (e.g., location and income in early phases). To improve cross-phase comparability for sex, we harmonized sex across versions by using \texttt{Biological Sex} when available and \texttt{ARCHIVED: Gender} otherwise. Descriptive distributions and missingness patterns for the U.S. and India cohorts are summarized in the Supplement.

\begin{figure}[htbp]
  \centering
  \includegraphics[width=0.96\linewidth]{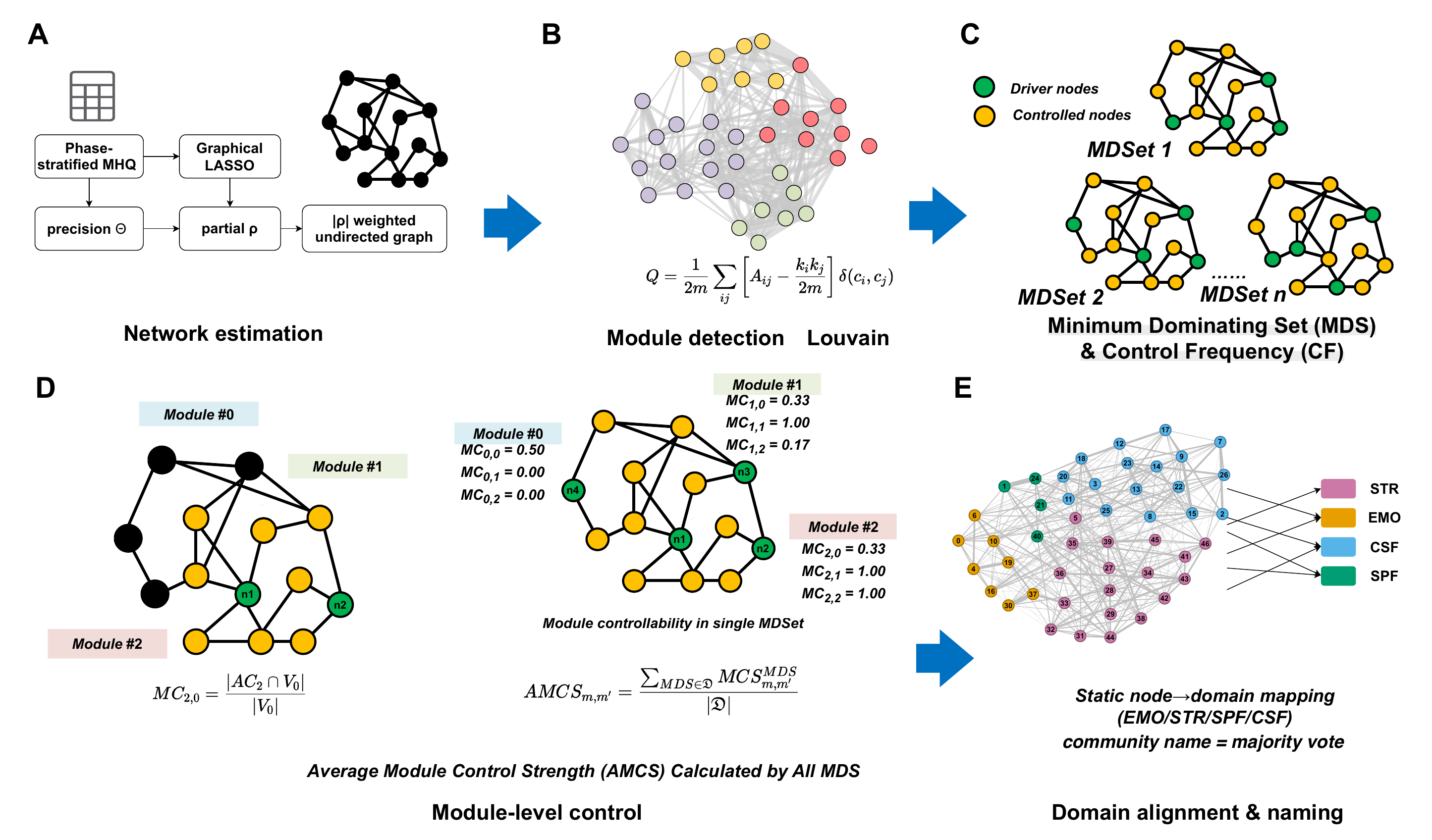}
  \caption{
\textbf{Analytical workflow for phase-stratified symptom-network and module-control analysis.}
(A) For each pandemic phase, Mental Health Quotient (MHQ) item scores (1--9) were used to estimate a Gaussian graphical model with graphical LASSO ($\alpha=0.40$). The precision matrix $\boldsymbol{\Theta}$ was transformed into partial correlations $\boldsymbol{\rho}$, and a weighted undirected network was constructed from $|\rho_{ij}|$, with edges retained whenever the penalized partial correlation was nonzero.
(B) Community structure was identified using the weighted Louvain algorithm ($\gamma=1.0$; fixed seed), and replicate partitions were aligned to the phase-specific full-sample partition by Hungarian matching on node--module overlap.
(C) On the unweighted support of the estimated graph, all minimum-cardinality dominating sets (MDSs) were enumerated exactly. Node control frequency, $\mathrm{CF}(v)$, was defined as the proportion of MDSs containing node $v$ and normalized to $[0,1]$.
(D) Module-level control was summarized by average control frequency (ACF) within each community and by average module control strength (AMCS), where $\mathrm{MCS}_{i\to j}$ denotes the proportion of nodes in community $C_j$ dominated by drivers in community $C_i$ across all MDSs; row-normalized AMCS defines a directed module control network (MCN).
(E) For cross-phase interpretation, nodes were mapped \emph{a priori} to four domains---emotional regulation (EMO), stress response (STR), self-perception and physiological function (SPF), and cognitive and social function (CSF). Communities were named by majority vote (\textit{Mixed} when purity was $<60\%$), and community-level control matrices were further aggregated into a $4\times4$ domain-level MCN.
}
  \label{fig:fig1}
\end{figure}

\subsection{MHQ items and preprocessing}
The MHQ is a 47-item instrument designed to capture a broad spectrum of emotional, cognitive, behavioural, and somatic symptoms on a common 1--9 response scale \cite{newson_assessment_2022,newson_mhq_2024}. Item wording, scoring conventions, and the correspondence between item IDs and item labels are provided in prior validation work and in Supplementary Table~S2. Because all items were recorded on the same metric, we did not apply additional rescaling or $z$-standardization.

Preprocessing followed GMP quality guidance and was applied separately within each phase. Records with abnormal completion times ($<7$ min or $>60$ min) were excluded to reduce the influence of rushed or inattentive responding. To further screen for invariant response patterns, we excluded respondents whose across-item response standard deviation was $<0.20$ on the 1--9 scale. No item-level imputation was performed. All network analyses therefore used listwise complete cases within each phase, such that each phase-specific covariance matrix was estimated from participants with complete data on all 47 items. For every bootstrap or case-dropping replicate (see below), covariance and precision matrices were re-estimated from the raw 1--9 scores using the same preprocessing pipeline, and partial correlations were obtained by normalizing the precision matrix to unit diagonal. Node IDs shown in figures correspond to MHQ item labels; the full mapping is provided in Supplementary Table~S2.

\subsection{Macro-level epidemiological and policy context}
To situate phase-specific youth symptom networks within the broader pandemic context, we assembled national epidemiological and policy indicators on a weekly grid. Daily confirmed cases and deaths in the United States were obtained from the New York Times COVID-19 dataset~\cite{times_coronavirus_2020} and aggregated to calendar weeks. For descriptive visualization, we used a 7-day rolling average of new confirmed cases per 100{,}000 population. For phase-level contextual summaries, we computed within-phase means and maxima of the national case series and of the Oxford COVID-19 Government Response Tracker (OxCGRT) composite stringency index. These descriptive summaries were used to annotate figures and to contextualize the phase-stratified analyses; they did not enter the estimation of symptom networks or module-control metrics.

For time-resolved exploratory analyses, we additionally aligned weekly MHQ trajectories with national epidemiological and policy time series. For each calendar week between 2020-01-01 and 2023-12-31, all available responses from the analytic population were averaged across respondents for each of the 47 MHQ items, yielding a weekly symptom series $\{x_t\}$ for each item. Weekly macro indicators comprised two epidemiological series (new confirmed cases and deaths, standardized as \texttt{new\_cases\_z} and \texttt{new\_deaths\_z}) and five OxCGRT policy indices \cite{hale_global_2021}: C1 (school closures), C3 (cancellation of public events), C6 (stay-at-home requirements), C7 (restrictions on internal movement), and C8 (international travel controls). These seven indicators were merged by calendar week to form the macro dataset $\{m_t\}$.

For each MHQ item and each macro indicator, we computed Spearman rank correlations between $\{x_t\}$ and lagged versions of $\{m_t\}$ over lags $\ell\in\{-12,-11,\ldots,0,\ldots,11,12\}$ weeks, using only weeks with complete data and requiring at least five overlapping observations. By convention, a positive lag ($\ell>0$) indicates that the macro indicator leads the symptom series. Primary interpretation focused on macro-leading lags from 0 to 12 weeks, consistent with the substantive view of epidemiological and policy variables as external contextual drivers; negative lags were retained as a sensitivity check for possible reverse ordering. For each item--macro pair, we recorded the lag $\ell^\ast$ that maximized the absolute correlation $|\rho|$, together with the corresponding $\rho$ and $p$-value.

Because 47 items were evaluated against seven macro indicators, raw $p$-values were adjusted across all item--macro pairs using the Benjamini--Hochberg false-discovery-rate (FDR) procedure~\cite{benjamini_controlling_1995}. These adjusted values were used for exploratory ranking only. Because the FDR procedure did not additionally correct for selecting the maximum $|\rho|$ over multiple lags, the lagged associations were interpreted as exploratory and pattern-generating rather than confirmatory. As a complementary robustness check, we also computed Granger-causality tests at 4- and 8-week lags after augmented Dickey--Fuller testing and first-differencing non-stationary series~\cite{dickey_distribution_1979,granger_investigating_1969}; these results are reported in the Supplement and were not used for primary inference.

\subsection{Network estimation and community detection}
All network analyses were conducted separately within each pandemic phase. For each phase, we estimated a Gaussian graphical model (GGM) \cite{lauritzen_graphical_1996} on the 47 MHQ items using the Graphical LASSO with a fixed $\ell_1$ penalty $\alpha=0.40$, chosen \emph{a priori} to maintain a moderately sparse representation across phases and to avoid data-adaptive tuning on the same observations used for subsequent control analyses \cite{friedman_sparse_2008,tibshirani_regression_1996}. From the estimated precision matrix $\boldsymbol{\Theta}$, we derived the partial-correlation matrix $\boldsymbol{\rho}$ as
\begin{equation}
\rho_{ij}=-\frac{\Theta_{ij}}{\sqrt{\Theta_{ii}\Theta_{jj}}}.
\end{equation}
We then constructed a weighted, undirected graph in which nodes corresponded to MHQ items and edge weights were the absolute partial correlations $|\rho_{ij}|$. An edge was present if and only if the corresponding partial correlation was nonzero under the $\ell_1$ penalty; no additional thresholding was applied. Node strength (weighted degree) was defined as the sum of $|\rho_{ij}|$ over all nonzero neighbours, following standard definitions for weighted networks \cite{barrat_architecture_2004,opsahl_node_2010}.

Communities were identified using the weighted Louvain algorithm with resolution parameter $\gamma=1.0$ and a fixed random seed (\texttt{random\_state}=42) for reproducibility \cite{blondel_fast_2008}. To enable valid comparison of module-level quantities across resamples and phases, replicate module labels were aligned to the corresponding phase-specific full-sample partition using Hungarian matching on the node--module overlap matrix \cite{kuhn_hungarian_1955}. When a replicate yielded fewer modules than the full-sample solution, unmatched replicate modules were assigned to the closest full-sample label by maximal overlap, and the resulting average module control strength matrices were embedded into the full-sample $K\times K$ space.

\subsection{Module control analysis}
Module-level control analysis followed the domination-based framework of Pan et al.\ \cite{pan_module_2024}. Control was defined on the unweighted support of the estimated graph; that is, an undirected edge between nodes $i$ and $j$ was retained whenever $|\rho_{ij}|>0$. On this support, we computed minimum dominating sets (MDSs), defined as smallest sets $D\subseteq V$ such that every node is either in $D$ or adjacent to at least one node in $D$ \cite{haynes_fundamentals_2013}. Although the MDS problem is NP-hard in general \cite{karp_reducibility_1972}, the modest graph size ($|V|=47$) made exact enumeration of all minimum-cardinality MDSs computationally feasible for each phase. We therefore enumerated the complete set of distinct minimum-cardinality MDS solutions for each phase-specific graph and computed CF/ACF/AMCS from the full solution set. Exact MDS counts and runtimes for all phases and both cohorts are reported in Supplementary Table~S8, and representative 3-node solutions for India Omicron are shown in Table~S10.

Let $f(v)$ denote the raw MDS participation rate of node $v$, defined as the proportion of enumerated minimum-cardinality MDSs that contain $v$. To obtain a within-graph relative control index comparable across nodes, we normalized this quantity by the maximum participation rate observed in the same graph:
\begin{equation}
\mathrm{CF}(v)=\frac{f(v)}{\max_{u\in V}f(u)}\in[0,1].
\end{equation}
Thus, the most frequently selected driver in a given phase has $\mathrm{CF}=1$. Throughout, \emph{control} refers to domination-based structural coverage in the estimated graph and should not be interpreted as a causal or intervention effect between symptoms.

For a community $C_k$ with node set $V_k$, the average control frequency (ACF) was defined as
\begin{equation}
\mathrm{ACF}_k=\frac{1}{|V_k|}\sum_{v\in V_k}\mathrm{CF}(v).
\label{eq:acf}
\end{equation}
To quantify directed control between communities, let $\mathrm{MCS}_{i\to j}$ denote the fraction of nodes in community $C_j$ that are dominated by drivers from community $C_i$ when aggregating over all MDSs, including the case $i=j$. Row-normalization then yields the average module control strength
\begin{equation}
\mathrm{AMCS}_{i\to j}=
\frac{\mathrm{MCS}_{i\to j}}{\sum_{j'}\mathrm{MCS}_{i\to j'}},
\label{eq:amcs}
\end{equation}
and the resulting $K\times K$ matrix $[\mathrm{AMCS}_{i\to j}]$ defines a directed module control network (MCN) for each phase \cite{pan_module_2024}. We used the unweighted support of $|\rho|$ because domination is a reachability notion: a neighbour connected by a negative partial correlation still renders a node structurally covered.

\subsection{Domain alignment and cross-phase control summaries}
To facilitate interpretation of data-driven communities, we used a static node-to-domain mapping that assigned each MHQ item to one of four pre-specified domains: emotional regulation (EMO), stress response (STR), self-perception \& physiological function (SPF), and cognitive \& social function (CSF). This lookup table was fixed \emph{a priori} across all phases and resamples, thereby avoiding circularity with fitted control metrics. The mapping yielded the following domain counts: STR = 18, CSF = 12, SPF = 9, and EMO = 8.

For each phase and each resample, communities obtained from the Louvain partition were named by majority vote over the domain labels of their member items. A purity threshold of 60\% was used; communities that did not reach this threshold were labelled \textit{Mixed} using the two most represented domains. Because the number of detected communities could differ from four, multiple communities could map to the same domain and some domains could be absent in a given phase. To ensure cross-phase comparability, community-level quantities were therefore embedded into a fixed four-domain space. Domain-level ACF was defined as the mean normalized $\mathrm{CF}$ over all items assigned to the same domain. Domain-to-domain AMCS was computed by summing community-level $\mathrm{MCS}_{i\to j}$ across communities mapped to the corresponding source and target domains, followed by row-normalization within the source domain, yielding a $4\times4$ domain-level MCN for each phase. In the First-wave phase, community detection occasionally split SPF items into two communities; for visualization we label the smaller SPF-dominant community as `PHY' (a physiological subcluster), but it is treated as part of SPF when aggregating results into the fixed four-domain (EMO/STR/SPF/CSF) space.

To characterize whether node-level control was persistent or phase-sensitive across the pandemic, we summarized each node's control trajectory over the five phases ($T=5$) using its mean normalized control frequency and coefficient of variation:
\begin{equation}
\overline{\mathrm{CF}}(v)=\frac{1}{T}\sum_{t=1}^{T}\mathrm{CF}_t(v), \qquad
\mathrm{CV}(v)=\frac{\mathrm{sd}\{\mathrm{CF}_t(v)\}_{t=1}^{T}}{\mathrm{mean}\{\mathrm{CF}_t(v)\}_{t=1}^{T}}.
\end{equation}
Nodes were then located in the $\overline{\mathrm{CF}}$--$\mathrm{CV}$ plane using median-based reference lines. Let
\[
m_{\mathrm{CF}}=\mathrm{median}_v\,\overline{\mathrm{CF}}(v), \qquad
m_{\mathrm{CV}}=\mathrm{median}_v\,\mathrm{CV}(v).
\]
Using tolerance bands $\delta_{\mathrm{CF}}=0.02$ and $\delta_{\mathrm{CV}}=0.05$, we classified nodes as \emph{Backbone} if they showed high average control and low variability,
\[
\overline{\mathrm{CF}}(v)\ge m_{\mathrm{CF}}+\delta_{\mathrm{CF}}
\quad \text{and} \quad
\mathrm{CV}(v)\le m_{\mathrm{CV}}-\delta_{\mathrm{CV}},
\]
and as \emph{Liaison} if they showed high average control and high variability,
\[
\overline{\mathrm{CF}}(v)\ge m_{\mathrm{CF}}+\delta_{\mathrm{CF}}
\quad \text{and} \quad
\mathrm{CV}(v)\ge m_{\mathrm{CV}}+\delta_{\mathrm{CV}}.
\]
All remaining nodes were retained as unlabeled reference points. The qualitative conclusions were unchanged under $\pm 0.01$ perturbations of the tolerance-band widths.

\subsection{Robustness and stability analysis}
We assessed robustness separately within each phase using two resampling schemes. First, we performed equal-size nonparametric bootstrap with replacement (1{,}000 resamples per phase), where each resample drew $k=n_{\text{phase}}$ participants \cite{efron_bootstrap_1979,efron_introduction_1994}. Second, we conducted case-dropping subsampling without replacement at retain ratios of 0.90, 0.80, 0.70, 0.60, and 0.50, again using 1{,}000 subsamples at each retain ratio. For every replicate, we re-estimated the GGM, re-ran community detection with the same fixed seed, enumerated MDSs, and recomputed three target metrics: node strength (over nodes), ACF (over communities), and AMCS (vectorized in row-major order from the $K\times K$ matrix). All replicate modules were label-aligned to the corresponding phase-specific full-sample partition before aggregation.

Stability was quantified as the Pearson correlation between each replicate metric and its full-sample counterpart: elementwise over nodes for strength, over communities for ACF, and over the vectorized AMCS matrix for MCN structure. We summarized drop-robustness using a control-stability (CS) coefficient, defined as the largest drop proportion (i.e., $1-\text{retain}$) for which the 5th percentile of the replicate--baseline correlation distribution remained at least 0.70, following established recommendations for network-metric stability \cite{epskamp_estimating_2018}.

\section{Results}\label{sec:results}

\subsection{Conserved modular organization across pandemic phases}
We first examined whether youth symptom networks retained a comparable mesoscale organization across pandemic phases, thereby providing a stable scaffold for phase-wise comparisons of controllability. Across the five phases, community detection consistently identified four to five communities (Fig.~2A), indicating that the network did not fragment into qualitatively different architectures as the pandemic evolved. Although the precise assignment of individual nodes varied at community boundaries, the overall partition remained highly similar across phases.

Domain-level summaries of classical network metrics (Fig.~2B) showed reproducible contrasts among functional domains, suggesting that broad differences in the structural roles of emotional, stress-related, cognitive/social, and self-perception/physiological symptoms were preserved over time. At the node level, trajectories remained heterogeneous (Fig.~2C): no single set of nodes emerged as a universally dominant hub class across all phases, and local patterns of prominence changed over time. Taken together, these findings indicate that the youth symptom system preserved a relatively stable modular scaffold despite the substantial changes in epidemiological conditions and policy regimes occurring between 2020 and 2023.

\begin{figure}[htbp]
  \centering
  \includegraphics[width=\linewidth]{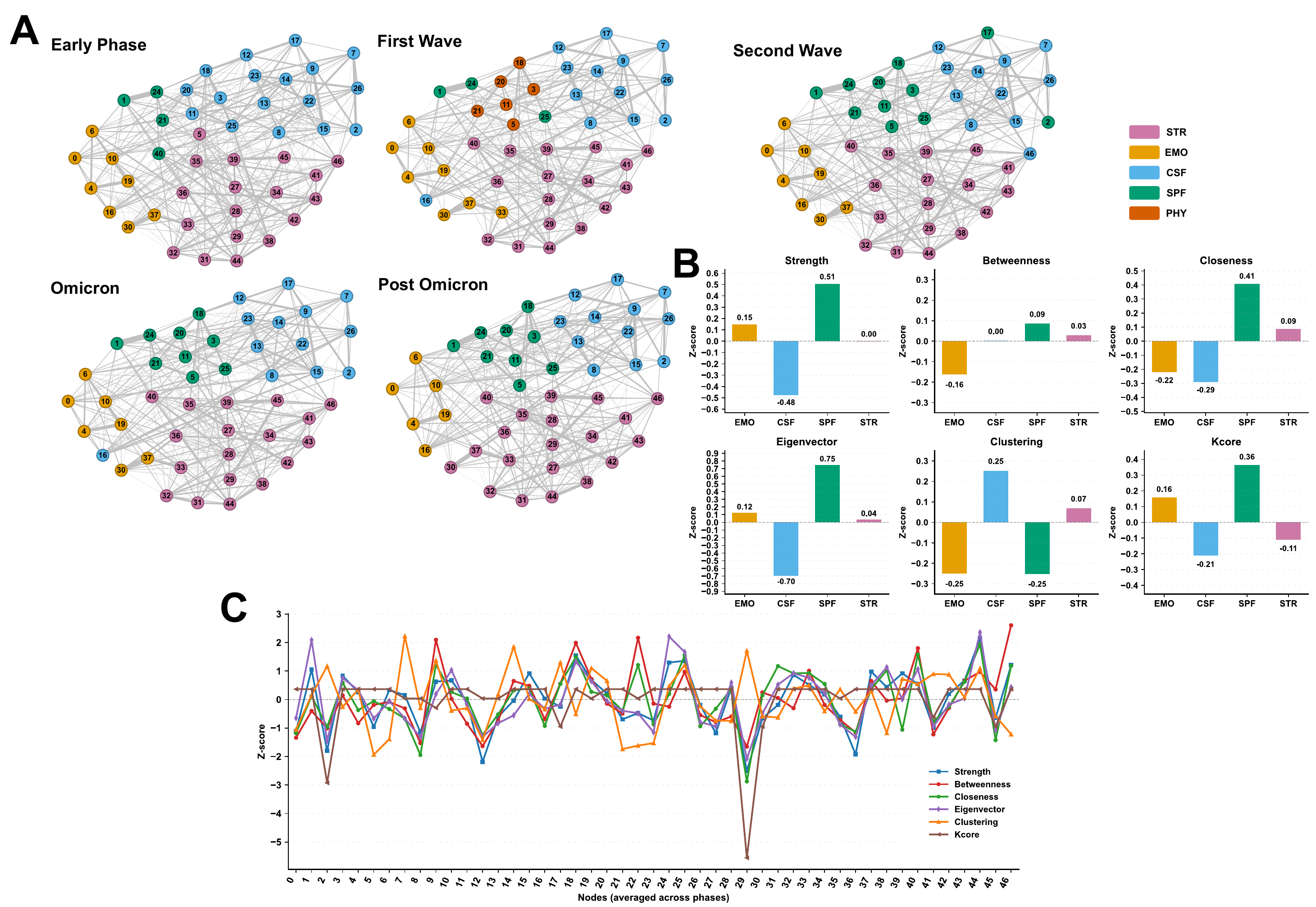}
  \caption{Conserved modular organization across pandemic phases. (A) Phase-specific symptom networks for the Early, First Wave, Second Wave, Omicron and Post-Omicron periods. Nodes are shown according to the detected community structure and aligned to a unified four-domain interpretation comprising stress response (STR), emotional regulation (EMO), cognitive and social function (CSF), and self-perception and physiological function (SPF). (B) Domain-level summaries of classical network topology, shown as phase-averaged standardized values for strength, betweenness, closeness, eigenvector centrality, clustering coefficient and k-core. (C) Node-level standardized profiles of the same network metrics, averaged across phases, illustrating heterogeneous local prominence despite broad conservation of the mesoscale scaffold.}
  \label{fig:fig2}
\end{figure}
\FloatBarrier

\paragraph{Cross-national replication of the mesoscale scaffold in India.}
To evaluate whether the conserved community scaffold observed in the U.S. cohort generalizes beyond a single country, we replicated the full pipeline in an independent India cohort (ages 18--24) using identical phase definitions, preprocessing, and estimation settings. Louvain community detection again yielded a small number of mesoscale modules per phase (India: $K=4$ across phases), closely matching the compact module regime observed in the U.S. (U.S.: $K=4$ in four phases and $K=5$ in First Wave). When aligning phase-specific partitions between countries by maximum-overlap matching, cross-national module correspondence was moderate-to-high across all phases (mean Jaccard overlap $\approx 0.52$--$0.72$) and became near-identical in the post-Omicron phase (mean Jaccard $\approx 0.94$; Supplementary Table~S6). Domain annotations based on the fixed node-to-domain mapping revealed high-purity modules for the primary domains (EMO/STR/CSF/SPF), with the First-wave phase occasionally showing a small SPF-dominant physiological subcluster (PHY) as an additional community label, supporting a conserved mesoscale organization across countries despite differences in cohort composition (Supplementary Table~S7).

\subsection{Redistribution of intermodule control across phases}

We next examined whether controllability was redistributed across domains on top of this stable scaffold, and the results indicated a clear redistribution of influence. While the modular partition remained broadly similar across phases, the module control network (MCN) showed clear temporal reweighting of influence (Fig.~3A). In the Early and First-wave phases, outgoing control was concentrated in stress-related circuitry: the STR domain displayed the highest out-degree and exerted strong influence over EMO and CSF, yielding a distinctly stress-dominated control configuration.

This pattern shifted in later phases. By the Omicron and Post-Omicron periods, control became less concentrated and more evenly shared across domains. In particular, SPF and CSF contributed more substantially to outgoing influence alongside STR, consistent with a transition from stress-dominated organization to a broader multi-module governance pattern. Domain-level ACF trajectories (Fig.~3B) and the corresponding in-degree and out-degree summaries (Fig.~3C--D) supported the same interpretation, showing that both control emission and control reception were redistributed over time.

These findings indicate that, despite the preservation of a relatively stable mesoscale modular architecture, the distribution of effective control across functional subsystems exhibited systematic reconfiguration over the course of the pandemic.

\begin{figure}[!t]
  \centering
  \includegraphics[width=0.95\linewidth]{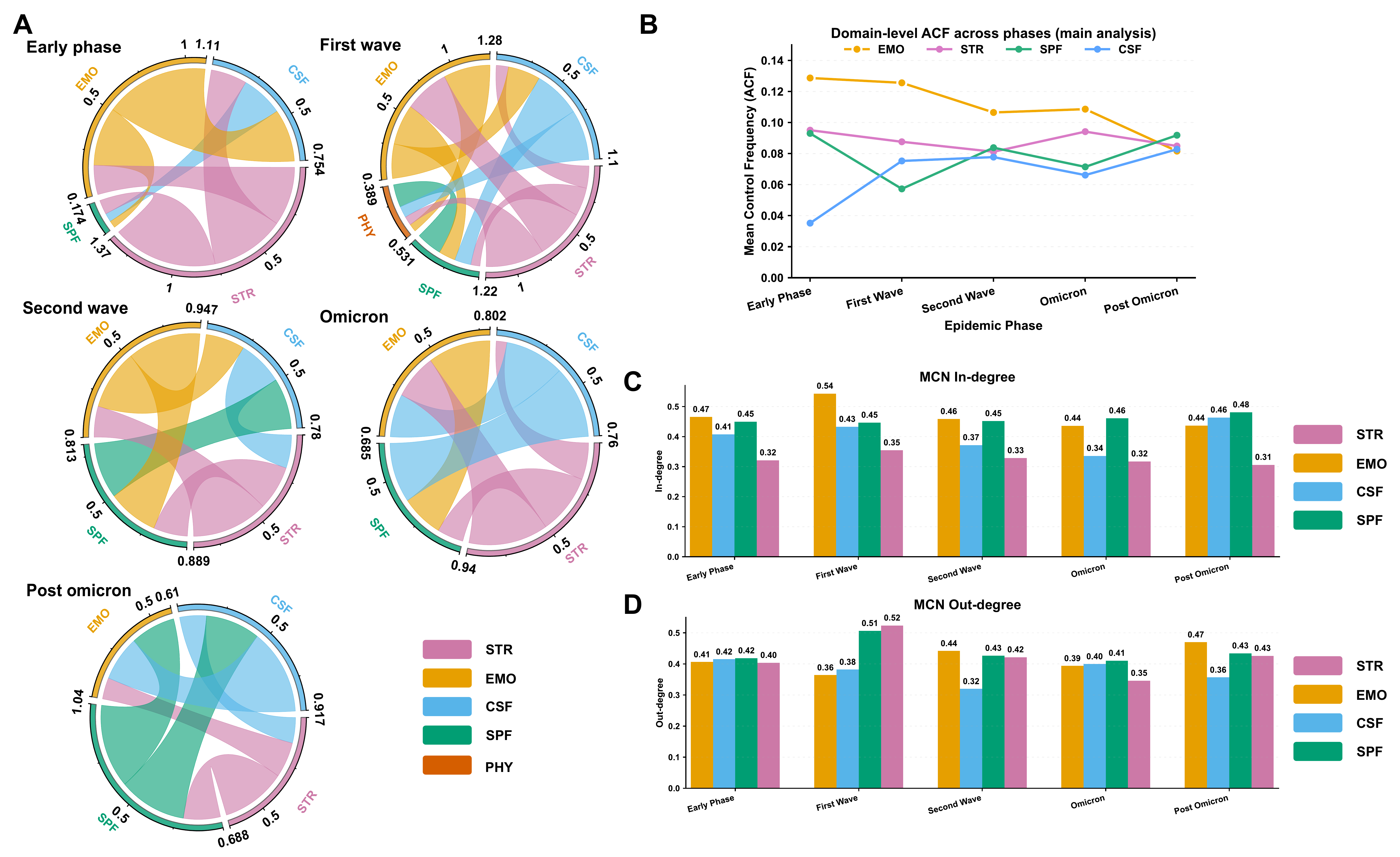}
  \captionsetup{font=footnotesize}
  \caption{Redistribution of intermodule control across pandemic phases. (A) Domain-level module control networks (MCNs) for each pandemic phase, derived from exact enumeration of all minimum dominating sets. Circular sectors represent domain-labeled modules; in the First-wave phase, a small SPF-dominant physiological subcluster (PHY; 6 nodes) is shown as a separate module label for visualization, while domain-level aggregation remains in the fixed four-domain (EMO/STR/SPF/CSF) space. Ribbon widths indicate normalized average module control strength (AMCS) from a source domain to a target domain. (B) Domain-level average control frequency (ACF) across phases, summarizing the normalized participation of symptoms from each domain in minimum dominating sets. (C) Domain-level MCN in-degree, quantifying the amount of incoming control received by each domain. (D) Domain-level MCN out-degree, quantifying the amount of outgoing control exerted by each domain. Across phases, the control configuration shifts from a more stress-centered pattern in the early pandemic to a broader cross-domain allocation in later phases.}
  \label{fig:fig3}
\end{figure}
\captionsetup{font=normalsize}

\paragraph{Replication of domain-level control redistribution and sensitivity to sample size.}
To evaluate whether these mesoscale and control-redistribution patterns were specific to the U.S. cohort, we repeated the full pipeline in an independent India cohort drawn from the same GMP platform using identical phase definitions and preprocessing. Cross-national module correspondence within each phase was moderate to high (Supplementary Fig.~S6; Table~S6), and the phase-wise pattern of domain-level control concentration showed comparable trajectories, with downsampling analyses indicating that conclusions were not driven by India's larger sample sizes (Supplementary Fig.~S5; Tables~S8--S9).

\subsection{Persistent backbone nodes and phase-sensitive boundary mixing}
To refine the mesoscale picture, we examined phase-specific changes in node- and community-level control. A direct comparison of the Early and Post-Omicron networks (Fig.~4A) suggested greater mixing at module boundaries in the later phase, while still preserving a subset of consistently influential controllers. This pattern implies that control migration was not driven by wholesale replacement of key nodes, but rather by reweighting of influence pathways around a persistent backbone.

The CF--CV dispersion plot (Fig.~4B) further clarified this distinction. Nodes with high mean control frequency and low cross-phase variability can be interpreted as backbone controllers, whereas nodes with high mean control frequency but greater variability function more like liaison nodes whose importance depends on phase-specific context. Several items repeatedly appeared among the most persistent high-control nodes, including ID~28 (Fear \& Anxiety), ID~37 (Anger \& Irritability), ID~44 (Feelings of Sadness, Distress or Hopelessness), ID~20 (Focus \& Concentration), and ID~10 (Emotional Resilience). These nodes combined relatively high average control frequency with comparatively low variability across phases, consistent with a stable core of influential symptoms that remained active throughout the pandemic.

The Sankey summary of cross-phase flows (Fig.~4C) further illustrated how control was reallocated across domains over time. Rather than indicating abrupt restructuring, the flows suggest a gradual redistribution of control away from an early concentration in stress-related processes toward a more distributed cross-domain pattern. Together, these findings indicate that phase-specific control migration was driven primarily by changes in inter-module allocation and boundary mixing, while a core set of backbone controllers persisted throughout the study period.
\begin{figure}[htbp]
  \centering
  \includegraphics[width=\linewidth]{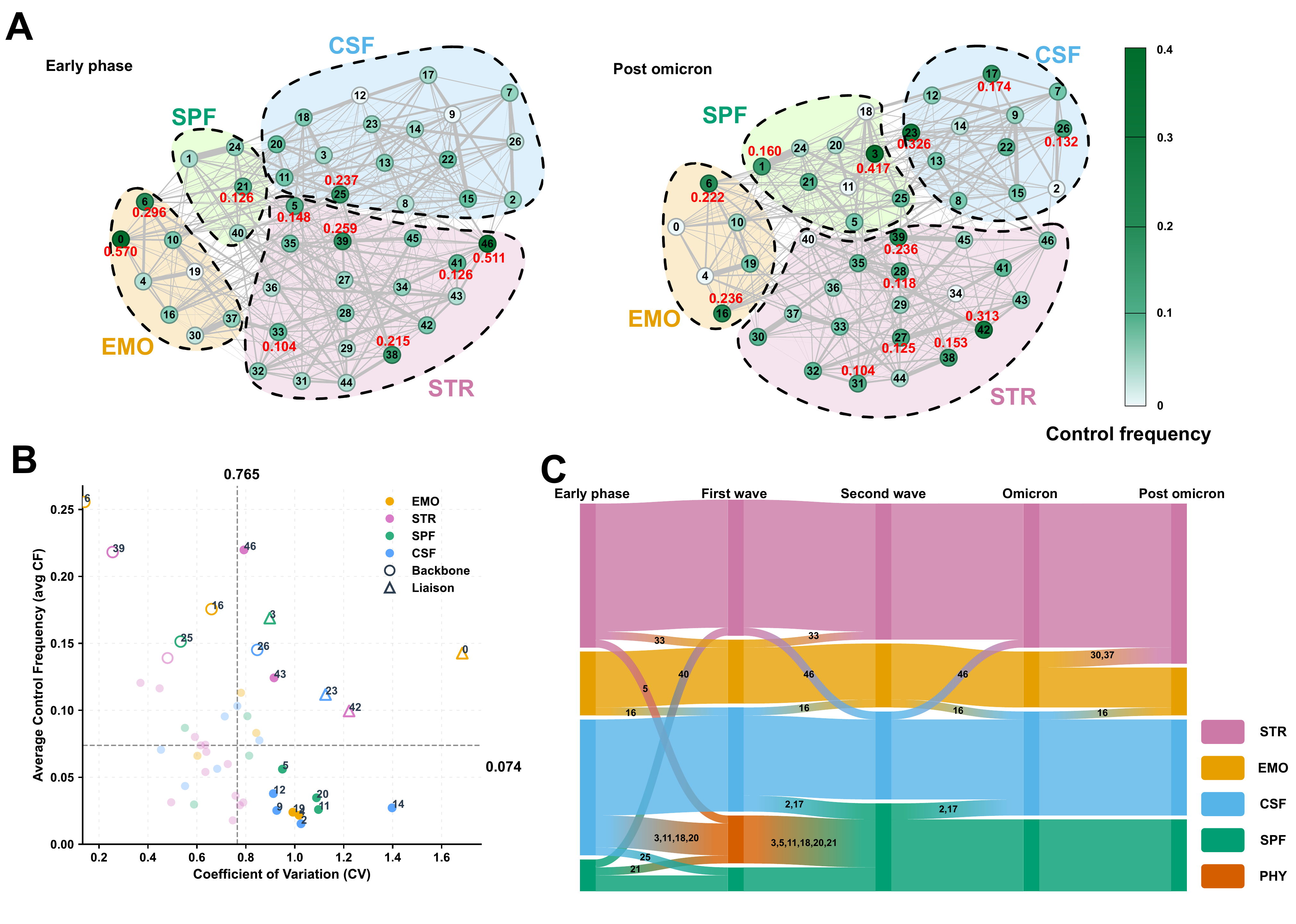}
  \caption{Persistent backbone nodes and phase-sensitive boundary mixing. (A) Comparison of the Early and Post-Omicron symptom networks, with nodes grouped by domain and shaded according to normalized control frequency. The later phase shows greater boundary mixing while retaining a subset of consistently influential high-control symptoms. (B) Mean control frequency (CF) plotted against the coefficient of variation (CV) across the five pandemic phases for each symptom. Median-based reference lines and tolerance bands are used to distinguish backbone nodes, defined by high average control and low variability, from liaison nodes, defined by high average control and high variability. (C) Sankey summary of cross-phase control reallocation across domains. Flows indicate how high-control symptoms are redistributed over time without evidence of wholesale restructuring.}
  \label{fig:fig4}
\end{figure}

\subsection{Macro-level policy context and youth symptom dynamics}
We then examined whether macro-level epidemiological and policy conditions aligned with youth symptom trajectories, thereby providing external context for the observed phase-wise reweighting of control. Figure~6A summarizes the broader U.S. pandemic environment from 2020 to 2023. Weekly case counts rose sharply in the initial phase, remained elevated across subsequent waves, peaked again during Omicron, and declined thereafter. Policy indicators showed a different temporal pattern: school closures, cancellations of public events, stay-at-home requirements, and movement restrictions were tightened rapidly early in the pandemic and then relaxed only unevenly over time. Notably, some restrictions---especially international travel controls---remained high even when case counts had fallen, suggesting a partial decoupling between epidemiological pressure and policy stringency in later phases.

Exploratory lagged Spearman analyses revealed several interpretable alignments between macro conditions and weekly MHQ item trajectories (Fig.~6B). Fear \& Anxiety was positively associated with domestic movement restrictions (C7), with the strongest association at an approximately 11-week macro-leading lag, suggesting that sustained domestic restriction preceded elevated anxiety. Decision-making \& Risk-taking showed a similarly delayed but negative association with C7, consistent with lower self-regulatory functioning following prolonged restriction. Emotional Resilience displayed a short-lag positive association with international travel controls (C8), peaking at roughly one week, whereas Physical Intimacy showed a longer-lag negative association with the same policy indicator, with an optimal lag of approximately six weeks.

The broader heatmap (Fig.~6C) generalized these patterns across all tested symptom--macro pairs. The strongest associations, ranked by FDR-adjusted significance, were more often linked to policy variables than to epidemiological indicators. Domestic movement restrictions and home-confinement policies (C6--C7) were overrepresented among the strongest links, especially for cognitive and self-regulatory items such as Decision-making, Ability to Learn, Selective Attention, Speech \& Language, and Energy Level, typically at lags of 8--12 weeks. International travel controls (C8) showed pronounced associations with Emotional Resilience, Self Worth, Physical Intimacy, and Relationships, often at shorter lags. By contrast, case counts and deaths appeared less frequently among the top associations and showed weaker, more diffuse relationships overall.

These exploratory analyses do not establish causality, but they do suggest that youth symptom dynamics were more closely aligned with structured policy environments than with infection counts alone. At minimum, they identify plausible macro time scales---ranging from weeks to a few months---over which external policy regimes may have contributed to the phase-specific reweighting of module-level control.
\begin{figure}[htbp]
  \centering
  \includegraphics[width=\linewidth]{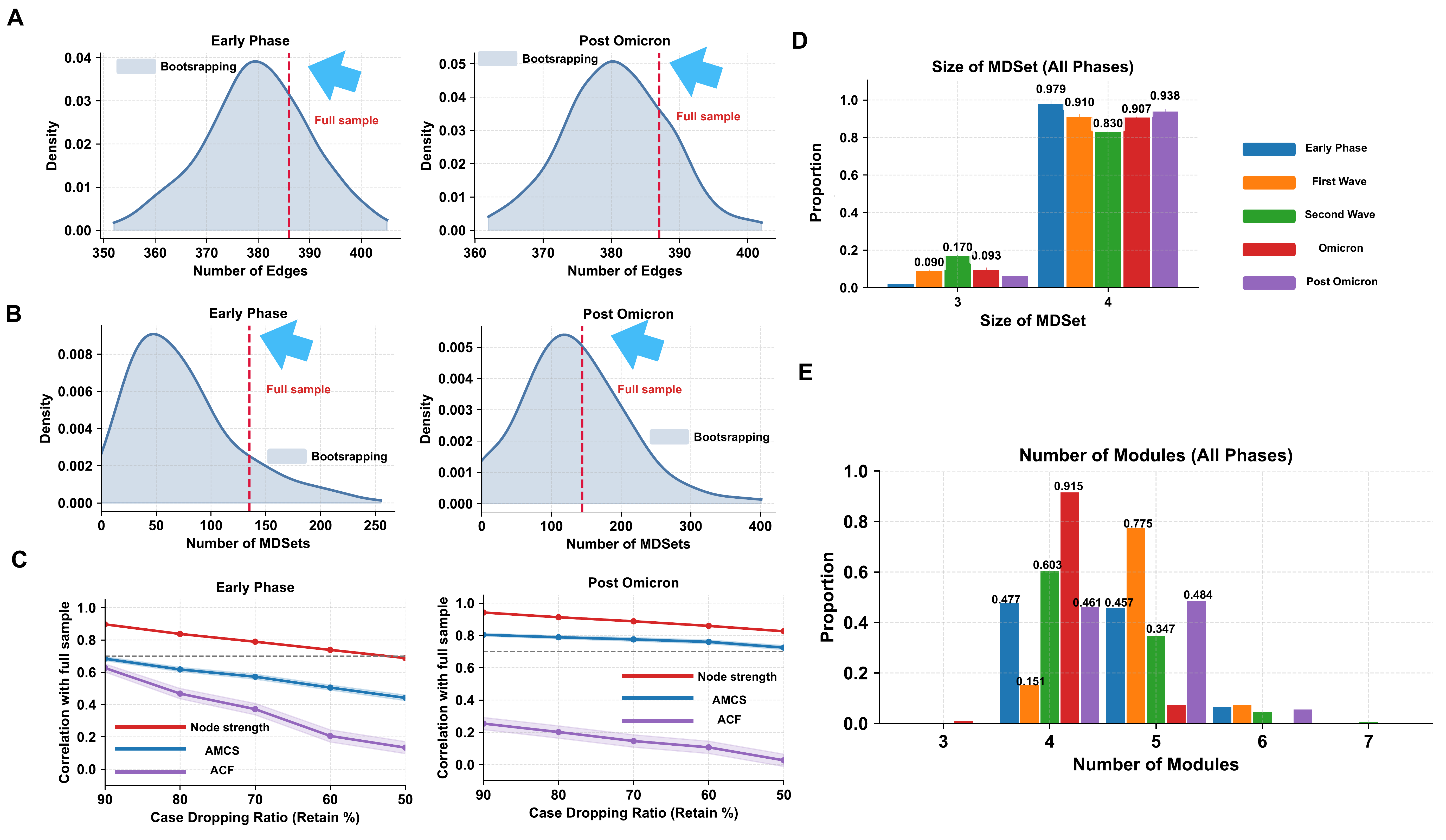}
  \caption{Robustness of network and control metrics. (A) Equal-size bootstrap distributions of the number of edges in the Early and Post-Omicron networks. Dashed red lines indicate the corresponding full-sample estimates. (B) Equal-size bootstrap distributions of the number of minimum dominating sets in the Early and Post-Omicron networks, with dashed red lines marking the full-sample values. (C) Case-dropping stability curves for node strength, average module control strength (AMCS), and average control frequency (ACF) in the Early and Post-Omicron phases, showing replicate-to-baseline correlations as the retained sample proportion decreases. (D) Phase-wise distributions of minimum dominating-set size across resamples. (E) Phase-wise distributions of the number of detected modules across resamples.}
  \label{fig:fig5}
\end{figure}

\subsection{Robustness of network and control metrics}
Finally, we assessed the robustness of key network and control metrics to resampling and case loss. Equal-size bootstrap analyses confirmed that the full-sample edge counts and minimum dominating set cardinalities fell within the 95\% resampling ranges in both the Early and Post-Omicron phases (Fig.~5A--B, D--E), indicating that the basic sparsity and domination structure of the estimated networks were not driven by anomalous samples.

Case-dropping analyses revealed the expected monotonic decline in replicate-to-baseline correlations as retain ratios decreased, but also showed a clear separation among metrics (Fig.~5C). Node strength was the most stable measure, AMCS showed moderate stability, and ACF was consistently the most sensitive to case loss. At 60\% retention, the phase-wise mean correlations with the full-sample metrics were 0.838 for node strength (range: 0.739--0.889), 0.737 for AMCS (range: 0.505--0.895), and 0.339 for ACF (range: 0.107--0.635). Thus, even under substantial sample reduction, strength and AMCS retained moderate to high correspondence with the full-sample estimates, whereas ACF degraded markedly.

The control-stability coefficients further summarized this gradient in robustness. Across phases, node strength remained acceptable under relatively large case loss, with median CS values around 50\%. AMCS was less robust but still retained limited drop tolerance in some phases, with CS values ranging from 0\% to 20\%. By contrast, ACF showed essentially no stable drop tolerance. Overall, these results indicate that node strength and module-to-module control are the most reliable metrics for phase-wise comparison in the present framework, whereas ACF should be interpreted more cautiously as a descriptive summary rather than a primary quantitative endpoint.

\begin{figure}[htbp]
  \centering
  \includegraphics[width=\linewidth]{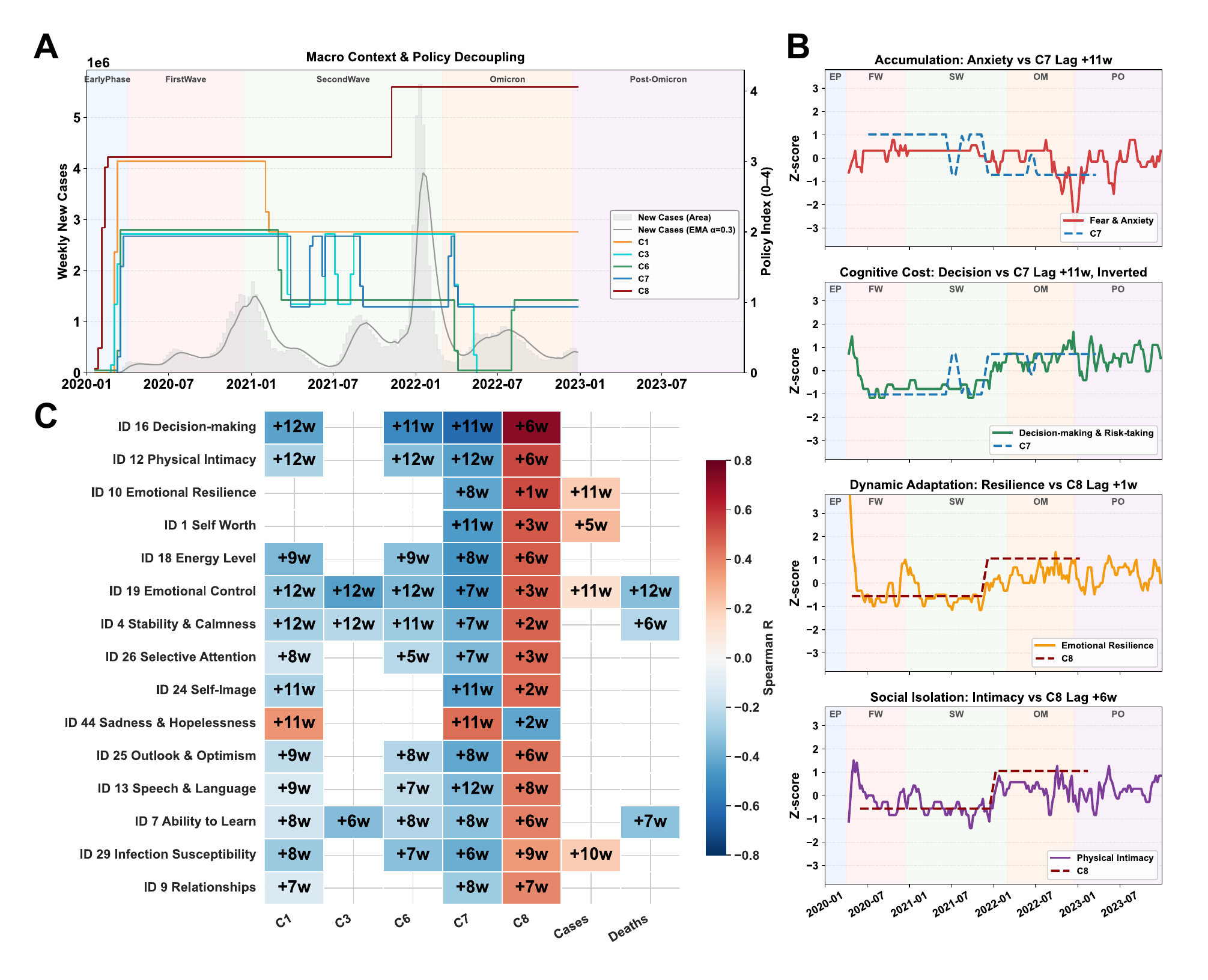}
  \caption{Exploratory weekly symptom trajectories in external macro context. (A) Weekly U.S. COVID-19 case counts and selected external context indicators from 2020 to 2023, shown across the Early, First Wave, Second Wave, Omicron and Post-Omicron periods. External indicators include school closing (C1), cancellation of public events (C3), stay-at-home requirements (C6), restrictions on internal movement (C7), and international travel controls (C8). (B) Representative symptom–context pairs illustrating lagged alignments between standardized weekly symptom trajectories and external indicators. Examples shown are Fear \& Anxiety versus C7 at +11 weeks, Decision-making \& Risk-taking versus inverted C7 at +11 weeks, Emotional Resilience versus C8 at +1 week, and Physical Intimacy versus C8 at +6 weeks. Dashed lines indicate the corresponding contextual indicator, and positive lags denote context-leading associations.
(C) Heatmap of the strongest lagged Spearman associations between weekly symptom means and external indicators, including policy indices, cases and deaths. Cell color indicates the sign and magnitude of the strongest association identified across the tested lag window, and text labels indicate the corresponding lag in weeks. }
  \label{fig:fig6}
\end{figure}

\section{Discussion}\label{sec:discussion}

The present study examined repeated cross-sectional youth symptom networks across five phases of the COVID-19 pandemic using network psychometrics and module-level control analysis. Three main findings emerged. First, the youth symptom network showed broadly conserved mesoscale organization across phases: community detection consistently yielded four to five modules, and the overall partition remained similar despite temporal variation. Second, this structural continuity coexisted with a redistribution of domination-based intermodule control. Early phases were characterized by a more stress-centered control configuration, whereas later phases showed a broader cross-domain pattern involving stress, emotional, cognitive/social, and self-perception/physiological domains. Third, robustness analyses indicated that node strength was the most stable metric and that module-to-module control was more reliable for cross-phase comparison than average within-module control, which was substantially more sensitive to case loss. Taken together, these findings support a view of youth psychopathology under prolonged crisis as a system with relatively stable modular organization but shifting mesoscale control allocation, and they identify intermodule control as a comparatively informative and moderately robust feature for phase-wise comparison.

One of the most consequential implications of these findings is that large-scale social disruption need not dismantle the internal organization of youth symptom systems in order to alter how those systems behave. The persistence of a similar modular scaffold across all five phases suggests that the broad functional architecture of youth psychopathology retained a degree of organizational inertia even under the sustained pressures of the pandemic. This pattern is important because it implies that major external shocks may act less by generating entirely new symptom structures than by shifting the relative influence of existing subsystems. In this sense, the pandemic appears to have operated not as a force that rewired the network from scratch, but as a prolonged perturbation that redistributed control within a pre-existing modular system.

The shift from early stress dominance to later multi-module governance provides a plausible systems-level account of how this redistribution unfolded. In the Early and First-wave phases, stress-related circuitry exerted disproportionate outgoing control, especially over emotional and cognitive/social domains. This configuration is consistent with an acute-response regime in which uncertainty, disruption, and threat appraisal occupy a central organizing role. During the initial phase of the pandemic, young people were confronted with abrupt school and university closures, social isolation, family and financial instability, and widespread uncertainty regarding infection risk, all of which likely amplified stress-responsive processes. Under such conditions, it is reasonable that stress-related symptoms functioned as dominant controllers capable of shaping downstream emotional and cognitive experience.

By contrast, the later emergence of more evenly distributed control suggests that the youth symptom network did not simply recover to a pre-pandemic state. Instead, the Omicron and Post-Omicron phases appear to reflect a chronic adaptation regime in which multiple functional systems jointly mediated ongoing psychological burden. Even as formal restrictions were relaxed and parts of social life resumed, young people continued to navigate accumulated stress, altered social routines, educational disruption, and longer-term uncertainty about relationships, identity, and future planning. Within this broader context, the growing contribution of SPF and CSF to outgoing control is consistent with a more diffuse pattern of dysregulation in which bodily self-perception, energy, cognitive efficiency, and social functioning become increasingly entangled with affective symptoms. Conceptually, this transition from stress-dominated control to multi-module governance suggests that prolonged crises may initially concentrate influence in acute threat-processing systems and only later redistribute control toward a broader network of adaptive and maladaptive functioning.

The node-level analyses refine this interpretation by showing that phase-specific reorganization occurred around a persistent backbone rather than through wholesale replacement of influential symptoms. Several nodes, including Fear \& Anxiety, Anger \& Irritability, Feelings of Sadness, Distress or Hopelessness, Focus \& Concentration, and Emotional Resilience, remained relatively prominent across phases. These symptoms can be interpreted as stable controllers located near major interfaces between emotional, stress-related, and cognitive systems. At the same time, the increased boundary mixing observed in later phases and the distinction between backbone and liaison nodes suggest that control migration involved changes in the pathways through which influence propagated across modules. In other words, the stable backbone provided continuity, whereas liaison nodes and shifting inter-module relations provided flexibility. This distinction may be especially useful for future work because it implies that symptom networks under prolonged stress contain both persistent leverage points and phase-sensitive connectors whose importance depends on broader context.

The macro-level analyses help place these internal network changes within a wider social and policy environment. Although the exploratory lagged correlations do not support causal inference, they suggest that youth symptom dynamics tracked policy structure more closely than epidemiological burden alone. In particular, domestic movement restrictions and home-confinement policies showed stronger and more concentrated associations with several cognitive and self-regulatory symptoms than did case counts or deaths. This pattern is theoretically meaningful. Infection counts represent epidemiological pressure at the population level, but policies determine how that pressure is translated into daily lived experience: where people can go, whether they can gather, how schooling is organized, and how social relationships are constrained or mediated. For young people, whose developmental stage is especially sensitive to peer interaction, autonomy, mobility, and educational continuity, such policy-mediated constraints may be more proximally relevant to mental functioning than infection incidence per se. The present findings therefore support the view that the mental-health consequences of a public-health emergency are not reducible to pathogen exposure alone; rather, they emerge from the interaction between epidemiological threat and the social organization of everyday life.

This point also has broader implications for public mental-health strategy. If early acute phases are characterized by stress-dominated control, then interventions that target stress reactivity, uncertainty management, and anxiety regulation may exert disproportionate influence on the wider symptom system. Such interventions may include rapid-access psychological support, clear and credible risk communication, school-based stress-management programs, and family-focused resources that reduce uncertainty and perceived uncontrollability. However, the later pattern of multi-module governance suggests that interventions may need to broaden as crises become chronic. When control is distributed across stress, affective, cognitive/social, and self-perception/physiological domains, narrow symptom-specific approaches may have less leverage than integrative strategies addressing emotional regulation, cognitive functioning, fatigue, social reconnection, self-worth, and embodied wellbeing simultaneously. The network results therefore imply that phase-tailored intervention planning may be more appropriate than treating youth mental-health burden during prolonged crises as a static problem requiring a fixed response.

A second important contribution of the present study is methodological. The analysis shows that not all network-derived control metrics are equally robust. Node strength displayed the highest stability, and AMCS showed moderate but usable stability across case-dropping analyses, whereas ACF was markedly more sensitive to sample loss. This matters for both substantive interpretation and future study design. The relatively strong performance of node strength and AMCS suggests that these metrics may be suitable for comparative work across phases, cohorts, or settings, especially when sample sizes vary. By contrast, the instability of ACF indicates that module-level averages of node control frequency should be treated more cautiously and interpreted as descriptive summaries rather than primary inferential endpoints. More broadly, the findings underscore the value of embedding network-control analyses within explicit resampling frameworks before drawing substantive conclusions about temporal shifts or candidate intervention targets.

Several limitations should be acknowledged. First, the design is phase-stratified and cross-sectional rather than longitudinal, so the observed differences across phases cannot be interpreted as within-person trajectories or definitive causal effects of pandemic conditions. The results characterize repeated cross-sections of the same broad population, not the evolving symptom networks of the same individuals over time. Second, the data derive from an online self-report instrument and therefore remain vulnerable to selection effects, reporting bias, and measurement constraints common to digital mental-health surveys. Third, the macro-level analyses are exploratory and ecological. The lagged associations between policy indicators and symptom trajectories identify plausible time scales of alignment, but they do not establish that particular policies produced specific symptom changes. Fourth, the network model itself involves simplifying assumptions. Gaussian graphical models estimate conditional dependence structure rather than causal direction, Louvain partitions are resolution-dependent, and the control analysis is performed on the unweighted support of networks derived from absolute partial correlations, which necessarily discards sign information and finer distinctions in edge magnitude. Finally, the mapping from data-driven communities to predefined domains provides interpretive clarity but may also compress meaningful heterogeneity within modules.

These limitations point toward several directions for future work. Longitudinal or panel data would allow direct examination of within-person network reconfiguration and the temporal order of macro-level shocks, online behavior, and symptom dynamics. Multi-level designs linking geographic policy exposure, school context, digital media use, and individual symptom networks would be especially valuable for clarifying how macro environments are translated into psychological outcomes~\cite{xu_faces_2025}. Future studies could also examine whether similar patterns of stable modular scaffolds with migrating control are observed in other crises, such as natural disasters, economic shocks, or political violence, thereby testing whether the present findings reflect a general principle of psychopathology under prolonged social disruption. Finally, methodological extensions incorporating signed or weighted domination, alternative community-detection schemes, or dynamic latent-network hybrids may help refine the interpretation of controllability in symptom systems.

In conclusion, the present study suggests that youth psychopathology during the COVID-19 pandemic is best understood not as a sequence of unrelated symptom configurations, but as a modular system with enduring structural organization and shifting control architecture. Across pandemic phases, the network retained a relatively stable mesoscale scaffold while reallocating effective control from an early stress-dominated regime toward a broader, multi-module pattern. This combination of structural stability and control migration offers a useful systems-level framework for understanding how prolonged public-health crises reshape youth mental health, and it highlights node strength and module-to-module control as promising, comparatively robust markers for tracking that process across changing social environments.

\bibliographystyle{IEEEtran}  

\section*{Author Contributions}
X.Z.: Conceptualization, Methodology, Supervision, Writing -- original draft, Writing -- review \& editing; T.F.: Data curation, Formal analysis, Visualization, Writing -- original draft, Writing -- review \& editing.

\section*{Acknowledgments}
This work was supported by the National Natural Science Foundation of China [62176129], National Natural Science Foundation of China--Jiangsu Joint Fund [U24A20701], and the Hong Kong RGC Strategic Target Grant [grant number STG1/M-501/23-N].

\section*{Competing Interests}
The authors declare no competing interests.

\section*{Data Availability}
The data analysed in this study are openly available through the Global Mind Project (GMP) platform. Analysis code is available at \url{https://github.com/realfty/covid-mcn}. A companion website providing a concise overview of the study, together with links to the manuscript, supplementary information, and analysis resources, is available at \url{https://zxzok.github.io/covid-mcn-site/}.

\bibliography{covid-mcn}

\clearpage

\end{document}